\documentclass[a4paper,11pt]{article}
\pdfoutput=1 

\usepackage{jcappub} 

\usepackage[T1]{fontenc} 

\usepackage{graphicx}
\usepackage{caption}
\usepackage{subcaption}

\usepackage[table,usenames,dvipsnames]{xcolor}

\def\be{\begin{equation}} 
\def\ee{\end{equation}} 
\def\bea{\begin{eqnarray}} 
\def\eea{\end{eqnarray}} 

\title{\boldmath A Bayesian Framework for Cosmic String Searches in CMB Maps}

\author[a,b,c]{Razvan Ciuca,}
\author[a,b]{Oscar F. Hern\'andez}

\affiliation[a]{ 
Department of Physics, McGill University, \\
3600 rue University, Montr\'eal, QC, H3A 2T8, Canada}
\affiliation[b]{Marianopolis College, \\ 4873 Westmount Ave.,Westmount, QC H3Y 1X9, Canada}
\affiliation[c]{ 
School of Computer Science, McGill University, \\
3480 rue University, Montr\'eal, QC, H3A 0E9, Canada}

\emailAdd{razvan.ciuca@mail.mcgill.ca}
\emailAdd{oscarh@physics.mcgill.ca}

\abstract{There exists various proposals to detect cosmic strings from Cosmic Microwave Background (CMB) or 21 cm temperature maps. Current proposals do not aim to find the location of strings on sky maps, all of these approaches can be thought of as a statistic on a sky map. We propose a Bayesian interpretation of cosmic string detection and within that framework, we derive a connection between estimates of cosmic string locations and cosmic string tension $G\mu$. We use this Bayesian framework to develop a machine learning framework for detecting strings from sky maps and outline how to implement this framework with neural networks. 
The neural network we trained was able to detect and locate cosmic strings on noiseless CMB temperature map down to a string tension of $G\mu=5 \times10^{-9}$ and 
when analyzing a CMB temperature map that does not contain strings, the neural network gives a 0.95 probability that $G\mu\leq2.3\times10^{-9}$. 
}

\begin{document}
\maketitle
\flushbottom

\section{Introduction}
\label{sec:intro}
In recent years there has been a renewed interest in the possibility that
cosmic strings might contribute to the power spectrum of primordial
fluctuations.   Whereas cosmic strings cannot be 
the dominant source of the primordial fluctuations~\cite{Albrecht,Turok},  
they can still provide a secondary source of fluctuations.
Many inflationary scenarios constructed in the context of supergravity models lead to the 
formation of gauge theory cosmic strings at the end of 
the inflationary phase \cite{Jeannerot1,Jeannerot2}, and in a large class of brane
inflation models, inflation ends with the formation of a network of cosmic superstrings~\cite{Sarangi} which can be stabilized as macroscopic objects in certain string models~\cite{Copeland}.
Finally, cosmic superstrings are also a possible remnant
of an early Hagedorn phase of string gas cosmology \cite{Brandenberger:2008nx}.
In all of the above mentioned scenarios, both a scale-invariant spectrum of
adiabatic coherent perturbations and a sub-dominant contribution
of cosmic strings is predicted.  Thus searching for signatures of cosmic strings probes particle physics beyond the Standard Model,  
and since cosmic strings form during a symmetry breaking phase transition, constraining the string tension $\mu$ constrains the particle physics symmetry-breaking pattern. 
The gravitational effects of the string can be parametrized by the dimensionless constant $G\mu$, where $G$ is Newtons gravitational constant. For cosmic strings formed in Grand Unified models, $10^{-8}<G\mu<10^{-6}$ whereas cosmic superstrings have $10^{-12}<G\mu<10^{-6}$~\cite{Brandenberger:2014lta}. 

There exists various proposal for the detection of cosmic strings.  
A string moving between an observer and the surface of last scattering can lead to a step discontinuity in a Cosmic Microwave Background (CMB) temperature anisotropy map through the Gott-Kaiser-Stebbins (GKS) effect~\cite{Gott:1984ef,Kaiser:1984iv} of long strings. The search for this effect on WMAP data has lead to null detection and a limit of $G\mu< 1.5 \times 10^{-6}$~\cite{Lo:2005xt,Jeong:2004ut,Jeong:2006pi,Jeong:2010ft}. 
More stringent  constraints come from CMB anisotropies angular power spectrum. 
The CMB angular power spectrum on the WMAP data gives a constraint an order of magnitude stronger than that from the GKS effect~\cite{Dvorkin:2011aj},
with the strongest such constraint coming from the Planck  Collaboration~\cite{Ade:2013xla}.  They have placed an upper limit on the string tension for Nambu-Goto strings of $G\mu < 1.3 \times 10^{-7}$ at the 95\% confidence level. 

As long cosmic strings move they accrete matter into over-dense wakes which perturb the CMB light and the 21 cm line in particular. 
Future 21 cm redshift surveys could observe cosmic strings wakes through their distinctive shape in redshift space~\cite{Brandenberger:2010hn,Hernandez:2011ym,Hernandez:2012qs,Hernandez:2014jda,Brandenberger:2015ifa} or through the cross-correlation between CMB and 21-cm radiation from dark ages~\cite{Berndsen:2010xc}.  

A final way to detect cosmic strings is through the gravity waves emission by cosmic string loops and their effect on pulsar timing experiments. Through pulsar timing constraints, the North American Nanohertz Observatory for Gravitational Waves (NANOgrav) places limits of $G\mu < 3.3 \times 10^{-8}$ at the 95\% confidence level~\cite{Arzoumanian:2015liz}. Interferometer experiments such as the LIGO-Virgo Collaboration are also searching for the gravity wave background from loops~\cite{Aasi:2013vna,Abbott:2009ws}. Because these limits come from gravitational waves produced by cosmic string loops, and there is still no agreement on the 
loop distribution\footnote{
Whereas all modern simulations of Nambu-Goto cosmic string networks agree on the cosmic string loop distribution~\cite{Ringeval:2005kr, Lorenz:2010sm,Blanco-Pillado:2015ana}, the abelian Higgs simulations disagree with the Nambu-Goto results~\cite{Hindmarsh:2017qff}. 
}, 
these limits are considered less robust than those arising from long strings. For this reason the most stringent robust limits on the string tension continue to be those from Planck, $G\mu < 1.3 \times 10^{-7}$ at the 95\% confidence level. 

Much research has been done in recent years to find a more sensitive probe of cosmic strings in CMB and 21 cm intensity maps. Edge and shape detection algorithms such as the Canny algorithm~\cite{Cannypaper}, wavelets, and curvelets have been proposed and studied as alternatives to the power spectrum in looking for cosmic strings in these maps~\cite{Amsel:2007ki,Stewart:2008zq,Danos:2008fq,Hergt:2016xup}.  All of these works use the scale invariant analytic model of long strings described in~\cite{Perivolaropoulos:1992if} to simulate CMB temperature anisotropy maps. We also use in this model in our studies. 

References~\cite{Amsel:2007ki,Stewart:2008zq,Danos:2008fq}
used the Canny algorithm to look for GKS edges produced by long strings in the CMB temperature maps simulated with this model. 
They found more short edges in maps with strings which they interpreted as the disruption of long edges by Gaussian noise. We have reproduced the Canny algorithm analysis and find that the excess number of short edges do not correspond to the locations of cosmic strings and that the actual limits of detection provided by Canny are for string tension above $G\mu=10^{-6}$. This is comparable to the results that wavelet transforms give as shown in Table III of ref.~\cite{Hergt:2016xup}.
Motivated by a search for an algorithm which would allow the effects of long strings to stand out, ref.~\cite{Hergt:2016xup} also used curvelet transforms to analyse simulated CMB temparature maps, and found that strings could be detected down to a string tension of $G\mu=1.4 \times 10^{-7}$ at the 95\% confidence level if the contribution of noise was not more than $1.6\mu$K (see Table III in~\cite{Hergt:2016xup}). 

Both the Canny and the wavelet, curvelet analysis~\cite{Amsel:2007ki,Stewart:2008zq,Danos:2008fq,Hergt:2016xup} 
involve choosing a statistical test to measure the significance of the difference in number of edges between a sky map with strings and one without.  All of these approaches can be thought of as a statistic on a sky map and it remains unclear to what extent the choices that were made can be improved. In Canny changing the mean maximum gradient parameter changes the number and types of edges found. In the wavelet and curvelet analysis of reference~\cite{Hergt:2016xup} only one type of mother function was used, and as mentioned in that paper, ``there is a lot of room for exploring possible improvement of the detection algorithm by optimizing the mother functions chosen.''
Furthermore, the Canny proposals do not find the location of strings on sky maps. 

Here we propose a Bayesian interpretation of cosmic string detection and within that framework, we improve on both these shortcomings: 
1) we propose a  framework by which all of these approaches, as well as new ones, can be unified, developed and studied systematically, in which machine learning searches for an optimal detection approach;
2) we derive a connection between estimates of cosmic string locations and cosmic string tension $G\mu$. 
To simplify the discussion, we present our framework within the context of cosmic string detection in CMB temperature anisotropy maps through the GKS effect. However it could equally be applied to the detection of cosmic string wakes in 21 cm intensity maps~\cite{Brandenberger:2010hn,Hernandez:2011ym,Hernandez:2012qs,Hernandez:2014jda,Brandenberger:2015ifa}. 

In our view a position space analysis of the temperature maps is essential for improving detection thresholds. A single string produces a signal which is localised in position space, only the pixels in the immediate neighbourhood of the string are affected by its presence. However, in the Fourier representation, a string causes all the modes to change, this makes it extremely hard to identify specific pixels as belonging on a string. Identifying string locations is extremely desirable since it allows for further experiments concentrated on the region where the string was detected. 

\def\Gu{$G\mu$}
The approaches currently used in cosmic string detection~\cite{Amsel:2007ki,Stewart:2008zq,Danos:2008fq,Hergt:2016xup} take the entire sky map and estimate \Gu\ from it.
However in the machine learning approach we are proposing, we take the sky map and produce another map with probabilistic estimates of string locations. From this second map we then estimate \Gu. The second step used to estimate \Gu\ is a straightforward calculation whose input is the map of probabilities produced by the neural network. No additional machine learning is necessary to estimate \Gu\ from the map. By splitting our estimate of \Gu\ in this we way we gain two important advantages. First, our approach provides information on where cosmic strings are most likely located in the sky. Secondly, from a machine learning point of view the training set needed for a neural network to learn to produce a map of cosmic string locations is much smaller than what would be needed for this neural network to produce an estimate of \Gu\ from a sky map. This is because string effects are highly local, effectively producing one training sample per pixel, instead of one training sample per full map, hence machine learning approaches are much better suited to producing the string location map instead of directly estimating \Gu\ . Thus we have split up the problem into something machines can do well (produce a map of probabilities) and something we as humans can do well (calculate \Gu\ from a map of probabilities). 

Our paper is organized as follows. In section~\ref{bayesian} we 
present our Bayesian point of view which focuses on obtaining $P(G\mu \, | \, \delta_{sky})$, the probability distribution of the string tension $G\mu$ given a  sky map $\delta_{sky}$. We do this by outlining a procedure to estimate the string locations in the sky given a sky map, and deriving an expression for $P(G\mu \, | \, \delta_{sky})$, which uses the information about string locations.  
In section~\ref{Pstring} we discuss the probability distributions of string location maps. 
In section~\ref{machine} we explain how machine learning can be used to compute the string location map from a sky map and in the process optimize cosmic string detection in CMB sky maps. We then outline how the machine learning framework can be implemented with a neural network. The technical details of this neural network are presented in a following paper~\cite{nextpaper}.
In section~\ref{resultsStringLocations} we discuss our neural network's predictions for the location of strings and compare it to previous work, in particular the Canny algorithm. In section~\ref{resultsGmuPosterior} we present the neural network's predictions for the posterior probability distribution of the string tension $G\mu$ given a sky map. We find we can accurately predict the value of the string tension on simulated maps with a $G\mu$ as low as $4\times10^{-9}$. 
And if we analyze a map that contains no strings 
the neural network gives a 0.95 probability that the string tension is $G\mu<2.3\times10^{-9}$.  We give our conclusions and discuss future work in~\ref{theend}.

After the first version of this manuscript was submitted to the arXiv the work of ref.~\cite{McEwen:2016aob} was brought to our attention.  In that paper, a Bayesian inference of the posterior probability distribution of $G\mu$ is derived using wavelets as a statistic on the CMB sky map. In our framework we used a neural network, which we will argue is a more general approach. Their figure~11 is a plot of this probability distribution similar to our figures~\ref{fig:PGmu} and~\ref{becomesflat}.  They also provide an estimate of cosmic string temperature maps but not a direct link between string locations and their estimate for the posterior probability distribution of $G\mu$, as we will do below. 

\section{Bayesian Interpretation of String Detection}
\label{bayesian}

Let us identify what exactly it means for cosmic strings to exist in CMB anisotropy maps. Let $\delta_{sky}$ be the CMB temperature fluctuations actually observed in the sky and $P(\delta_{sky})$ to be the probability distribution of this sky data. Cosmic strings presumably have some effect on this distribution, i.e. the distributions $P(\delta_{sky}\, | \, \nexists~\text{cosmic strings})$ and $P(\delta_{sky} \, | \,  \exists~\text{cosmic strings})$ are not the same. 

The earlier attempts at detecting cosmic strings from CMB maps that were discussed in the introduction operate in a frequentist framework, where the null hypothesis is that cosmic strings do not exist. The overall structure of these approaches is to find a statistic $F: \mathbf{R}^{m\times m} \rightarrow \mathbf{R}$ for which $P(F(\delta_{sky})\, | \, \text{null hypothesis})$ is as low as possible for the lowest possible $G\mu$.
In contrast for us the problem of string detection from CMB temperature maps is to obtain the posterior probability distribution $P(G\mu \, | \,  \delta_{sky})$, i.e. estimate the probability distribution over $G\mu$ given the sky temperature map. 

However, simply obtaining the probability over $G\mu$ is not ideal: this gives us virtually no indication as to the location of the detected cosmic strings in the sky. It seems desirable to search for methods which are not only able to estimate the probability distribution above, but are also capable of providing locations of candidate strings, this would allow for other methods to search the regions of sky most likely to contain strings and independently confirm their existence, current methods of cosmic string detection from CMB maps such as Canny or wavelet analyses do not provide this information. 

In this section we aim to derive a link between estimates of string locations and estimates of $G\mu$, i.e. given a method for producing estimates of string locations, we wish to find a way to compute $P(G\mu\, | \,  \delta_{sky})$ . Obtaining such a link is desirable because obtaining estimates of string locations is actually easier than directly obtaining the posterior probability distribution over $G\mu$, assuming that we have some method of accurately simulating maps of given $G\mu$. Here "easier" is meant in terms of the number of map simulations required to learn to estimate $G\mu$ from the map, versus the number of simulations required to learn to estimate string locations. We can see that this is so by noticing that each simulated string temperature map provides two pieces of information, first, $G\mu$, and second, the pixel locations of every string. The latter contains much more information than the former, so in learning to identify string locations, it is much more data-efficient to learn to identify string locations rather than try to learn to estimate $G\mu$ directly from maps. 
 
We now derive an expression for $P(G\mu \, | \,  \delta_{sky})$ which uses information about the string locations.  Let $\xi$ be a map which indicates which pixels lie on a string. A map $\xi$ is associated with a CMB temperature map $\delta_{sky}$. If $(i,j) \in \textit{string}$ then $\xi_{i,j}=1$, otherwise $\xi_{i,j}=0$. Call the space of all such maps $\mathbf{\Xi}$.
The problem of providing the location of cosmic strings is then equivalent to estimating the probability distribution $P(\xi \, | \,  \delta_{sky}, G\mu)$, the probability that the map $\xi$ represents the true string locations given the sky temperature maps and the knowledge that the string tension is $G\mu$. In general this is an intractable function to evaluate since the space $\mathbf{\Xi}$ is so large. For simplicity we make the following assumption about the distribution:
\be\label{xiassumption}
	P(\xi \, | \,  \delta_{sky}, G\mu) = \prod_{(i,j)} (p_{i,j})^{\xi_{i,j}}(1-p_{i,j})^{1-\xi_{i,j}}
\ee
where $p_{i,j}$ is the probability that the pixels located at position $i,j$ on the map is on a string given the pixels values of the sky map, i.e. $p_{i,j}$ is a function of $\delta_{sky}$. Here we are modelling the probability distribution as a multiplication over all pixels of Bernoulli distributions. A major assumption here is the independence of each pixel given the entire sky map, this assumption encodes our belief that we should be able to decide whether a given pixel is on a string solely from the temperature map, without knowing anything about which other pixels are actually on a string. This assumption is likely not true since knowing whether neighbouring pixels are on a string should strongly influence our probability estimates. Nonetheless, the space $\mathbf{\Xi}$ is so large that we 
decided to 
make the above assumption if we are to make any progress at all. In the next section we will describe a machine learning approach for estimating the quantity above.

We now massage the above formula to obtain an expression for $P(G\mu\, | \, \delta_{sky})$ which is presented in eq.~\ref{PGsky}. First, by Baye's rule:
\be\label{Pijbayes}
	P(\xi \, | \,  \delta_{sky}, G\mu) = \frac{P(\delta_{sky}, G\mu \, | \,  \xi) \times P( \xi ) }{P(\delta_{sky}, G\mu )}
\ee
where
\[
P(\delta_{sky}, G\mu ) = P(\delta_{sky}) \times P(G\mu \, | \,  \delta_{sky})
\]
and 
\[
P(\delta_{sky}, G\mu \, | \,  \xi) = P(\delta_{sky}\, | \, \xi, G\mu) \times P(G\mu \, | \, \xi).
\]
Replacing these terms in equation \ref{Pijbayes} and isolating $P(G\mu \, | \, \delta_{sky})$ gives us
\[P(G\mu \, | \, \delta_{sky}) = \frac{P(\delta_{sky} \, | \, \xi, G\mu) \times P(G\mu \, | \, \xi) \times P(\xi)}{P(\delta_{sky})\times P(\xi \, | \, \delta_{sky}, G\mu)}~.\]
We assume that the string distribution itself does not provide us with information regarding the possible values of $G\mu$ so that $P(G\mu\, | \, \xi) = P(G\mu)$, hence
\bea\label{PGsky1}
P(G\mu \, | \, \delta_{sky}) = \frac{P(\delta_{sky} \, | \, \xi, G\mu) \times P(G\mu) \times P(\xi)}{P(\delta_{sky})\times P(\xi \, | \, \delta_{sky}, G\mu)}~.
\eea
Finally, summing over possible $\xi \in \mathbf{\Xi}$ we obtain the main result of this section:
\[
\sum_{\xi \in \mathbf{\Xi}} P(G\mu \, | \, \delta_{sky})  = 
\Bigg(
\frac{P(G\mu)}{P(\delta_{sky})} 
\Bigg)
\Bigg\{\sum_{\xi \in \mathbf{\Xi}}\frac{P(\delta_{sky} \, | \, \xi, G\mu) \times P(\xi)}{P(\xi \, | \, \delta_{sky}, G\mu)}\Bigg\} \, . 
\]
$\Xi$ contains $2^{N_{\rm pixel}}$ elements where $N_{\rm pixel}$ is the number of pixels in a map, hence
\bea\label{PGsky}
P(G\mu \, | \, \delta_{sky})  = \Big({1\over2}\Big)^{N_{\rm pixel}}\, 
\Bigg(
\frac{P(G\mu)}{P(\delta_{sky})} 
\Bigg)
\Bigg\{\sum_{\xi \in \mathbf{\Xi}}\frac{P(\delta_{sky} \, | \, \xi, G\mu) \times P(\xi)}{P(\xi \, | \, \delta_{sky}, G\mu)}\Bigg\}\, . 
\eea
Equation \ref{PGsky} is not the $\xi$-marginalized distribution of \ref{PGsky1}.  
The point of the summation is to ultimately remove the appearance of $P(\xi)$ by writing the terms in curly brackets as an expectation value which we can approximate by using a dataset of $\xi$ sampled from $P(\xi)$, without having explicit access to $P(\xi)$. The term $P(G\mu)$ is simply the prior probability over $G\mu$, 
and $P(\delta_{sky})$ is the prior probability of obtaining $\delta_{sky}$, which we calculate through its Fourier transform of $\delta_{sky}$ since $P(\delta_{sky}) = P(\tilde\delta_{sky})$.

In the last pages we have factored $P(G\mu \, | \,  \delta_{sky})$ in a way which makes use of $P(\xi \, | \,  \delta_{sky}, G\mu)$ and $P(\delta_{sky} \, | \,  \xi, G\mu)$. 
This is motivated by the desire to explicitly compute the string locations and to reformulate the problem in terms of quantities that can be more easily computed in a machine learning approach.
The  most troublesome term in eq.~\ref{PGsky} is $P(\xi \, | \,  \delta_{sky}, G\mu)$. We will  estimate it in Sec.~\ref{machine} using machine learning and simulation maps of the temperature fluctuations due to cosmic strings. To compute $P(\delta_{sky}\, | \, G\mu, \xi')$ and $P(\xi')$ we will use a dataset of generated string maps to estimate it using Monte Carlo sampling in Sec.~\ref{Pstring}. 
We continue our discussion and calculation of these probability distribution in the following section.

\section{Probability Distributions of String Maps}
\label{Pstring}

In order to calculate $P(G\mu\, | \,  \delta_{sky})$ from eq.~\ref{PGsky} we need to estimate
$
P(\xi) \, , \, P(\delta_{sky}\, | \, G\mu, \xi)$, 
and $P(\xi \, | \,  \delta_{sky}, G\mu)\, .
$
In this section we discuss how to calculate these first two probabilities.

For $P(\xi)$ notice that it always appears in an expression of the form
\bea
\sum_{\xi \in \mathbf{\Xi}}\frac{P(\delta_{sky} \, | \, \xi, G\mu) \times P(\xi)}{P(\xi \, | \, \delta_{sky}, G\mu)}
&=&
\Bigg\langle
\frac{P(\delta_{sky} \, | \, \xi, G\mu)}{P(\xi \, | \, \delta_{sky}, G\mu)}
\Bigg\rangle_{ \xi \sim P(\mathbf{\xi})}~
\nonumber\\
&\approx& \frac{1}{n} \sum_{i=1}^n  \frac{P(\delta_{sky} \, | \, \xi^i, G\mu)}{P(\xi^i \, | \, \delta_{sky}, G\mu)}
\label{replacecurly}
\eea
where each $\xi^i$ is sampled from the distribution $P(\xi)$. The true probability distribution of strings $P(\xi)$ is intractable to compute, however, we can generate string temperature maps from string evolution simulations to obtain samples from an approximation to the true distribution, the simulations of string evolution themselves do not compute samples from the true distribution, however, this is the best we can currently do, as simulations improve, so will the accuracy of this assumption. We use samples from the simulations to evaluate the term above. 

We model $\delta_{sky}$ as being composed of two different elements $\delta_{gauss}$ and $\delta_{string}$ such that
\be
\label{skyisgaussplusstring}
\delta_{sky} = \delta_{gauss} + G\mu\,  \delta_{string}  
\ee
$\delta_{gauss}$ is the standard $\Lambda$CDM cosmology CMB anisotropies that can be computed from the power spectrum.
The $\delta_{string}$ can be computed in the long string model of~ref.~\cite{Perivolaropoulos:1992if}. It is made up of the superposition of GKS temperature discontinuties of individual strings, each given by 
$
8 \pi G \mu \gamma_s [\hat{n}\cdot(\vec{v}_s\times\hat{e}_s)]
$,
where $\hat{n}$ is the direction of observation, $\vec{v}_s$ is the velocity of the string, $\hat{e}_s$ is its orientation, and $\gamma_s = 1/\sqrt{1-v_s^2/c^2}$.  A more detailed description of the procedure we used to simulate $\delta_{string}$ and $\delta_{sky}$ can be found in~\cite{Stewart:2008zq,Danos:2008fq}.

Thus we can further approximate $P(\delta_{sky}\, | \, \xi^i, G\mu )$ using simulated maps. 
In the following equations we will keep $\delta_{sky}$ fixed and consider conditional probabilities like $P(\delta_{sky} \, | \, \delta_{string})$ where we are conditioning on a string component which is not the same as the one making up $\delta_{sky}$. Likewise the conditioned $G\mu$ and $\xi$ are not necessarily the true ones associated with $\delta_{sky}$, hence inside integrals over $\delta_{string}$, $\delta_{sky}$ does not vary. Next, we write
\be\label{Psky1}
P(\delta_{sky}\, | \, \xi^i, G\mu) = \int d\delta_{string} P(\delta_{sky}\, | \, \xi^i, G\mu, \delta_{string}) \times P(\delta_{string}\, | \, \xi^i, G\mu)
\ee
Note that here neither the true $G\mu$, $\delta_{sky}$ or $\xi$ are known.
However, 
since $\delta_{sky}$ depends on $\xi$ only through $\delta_{string}$ we have:
\be
P(\delta_{sky}\, | \, \xi, G\mu, \delta_{string}) = P(\delta_{sky}\, | \, G\mu, \delta_{string})
\ee
Finally, if we are given the string component $\delta_{string}$ for certain, and we are also given $G\mu$ for certain, then the only uncertainty remaining is in the element $\delta_{sky} - G\mu\,  \delta_{string}$.  Thus $P(\delta_{sky}\, | \, G\mu, \delta_{string}) = P(\delta_{sky}-G\mu\, \delta_{string})$, which we can compute from its power spectrum.
It is  important to note that  $\delta_{sky}-G\mu\, \delta_{string}$ does not equal $\delta_{gauss}$ since the string tension $G\mu$ and the string map $\delta_{string}$ that we are conditioning on are not necessarily those in $\delta_{sky}$. 

Returning to the computation of $P(\delta_{sky} \, | \,  \xi^i, G\mu)$, eq.~\ref{Psky1} become:
\bea
P(\delta_{sky}\, | \, \xi^i, G\mu) &=& \int d\delta_{string} ~P(\delta_{sky}-G\mu \, \delta_{string} ) \times P(\delta_{string}\, | \, \xi^i, G\mu)
\nonumber\\
&=& 
 \Big\langle P(\delta_{sky} - G\mu\,  \delta_{string}  )
 \Big\rangle_{\delta_{string} \sim P(\delta_{string}\, | \, \xi^i, G\mu)} 
\nonumber\\
&\approx&  P (\delta_{sky} - G\mu\,  \delta_{string}^i )
\label{Psky|xiGmu}
\eea
where the last line come about because the distribution  $P(\delta_{string} \, | \, \xi^i, G\mu)$ is well approximated by a delta function centred around the true $\delta_{string}^i$ corresponding to $\xi^i$.

Using equations \ref{replacecurly} and \ref{Psky|xiGmu} we now rewrite eq.~\ref{PGsky} as
\be\label{PGskyBis}
P(G\mu \, | \, \delta_{sky})  \approx \Big({1\over2}\Big)^{N_{\rm pixel}}\, 
\Bigg(
\frac{P(G\mu)}{P(\delta_{sky})} 
\Bigg)
\Bigg\{
\frac{1}{n} \sum_{i=1}^n  \frac{ P (\delta_{sky} - G\mu\,  \delta_{string}^i ) }{ P(\xi^i \, | \, \delta_{sky}, G\mu)}
\Bigg\}
\ee
Eq.~\ref{PGskyBis} is the way we will calculate the probable value of the string tension. 
All that remains is to discuss how we evaluate $P(\xi^i \, | \, \delta_{sky}, G\mu)$ in the denominator. Recall that $P(\xi^i \, | \, \delta_{sky}, G\mu)$ estimates the location of cosmic strings in the sky map. We turn to its evaluation through machine learning in the next section.

\section{Machine Learning Optimization of String Detection}
\label{machine}

In equations~\ref{PGsky} and \ref{PGskyBis} we have derived an expressions for $P(G\mu \, | \,  \delta_{sky})$ in terms of other probability distributions, and in particular in terms of 
\be\label{Pijstring}
P(\xi\, | \,  \delta_{sky}, G\mu)
\ee
It is important to keep in mind that the string tension $G\mu$ we are conditioning on is not necessarily the string tension that is in the string contribution to $\delta_{sky}$. So for example, if we have a $\delta_{sky}$ with absolutely no string contribution, i.e. the real string tension is $0$, but we condition on a string tension of $10^{-6}$, then we would expect that any bit of noise is due to a string and $P(\xi\, | \,  \delta_{sky}, G\mu)$ will be higher for those $\xi$ with pixels on the highest noise part of the $\delta_{sky}$. In other words we will be imagining pixels as being on strings, even though they are not. On the other hand if the string tension is $\delta_{sky}$ is large but we condition on a $G\mu$ near $0$, then our $P(\xi\, | \,  \delta_{sky}, G\mu)$ will be close to uniformly distributed in ${\mathbf \Xi}$, the space of all string maps $\xi$. Pixels with signals that may appear to be due to strings will be assigned a low posterior probability of being on a string.

Calculating the distribution $P(\xi\, | \,  \delta_{sky}, G\mu)$ is an intractable problem. However we can use simulations to build a dataset of string temperature maps. We can then approximate \ref{Pijstring} by computationally sampling from the dataset along with machine learning. 
The idea is to parametrise the probability distribution by some function with a very large number of free parameters and then choose parameter values which give the best approximation to the distribution $P(\xi\, | \,  \delta_{sky}, G\mu)$.

Consider a very large number of parameters which we assemble into a parameter vector $\beta$. 
In the neural network example we will discuss below 
the vector $\beta$ has a dimension of order $10^5$. 
Let $P_\beta(\xi \, | \,  \delta_{sky}, G\mu)$ be the parametrised distribution that we will use to approximate $P(\xi \, | \,  \delta_{sky}, G\mu)$,
To find the best parameters $\beta$, we need to minimize some distance between the parametrised distribution $P_{\beta}(\xi \, | \,  \delta_{sky}, G\mu)$ and $P(\xi \, | \,  \delta_{sky}, G\mu)$.  To measure the distance between probability distributions we will use the Kullback-Leibler divergence, which we will define in eq.~\ref{KLdiv} below. 

We treat the probability of there being a string on pixel $(i,j)$ as a Bernoulli distribution with a pixel dependent probability of success  $p_{i,j}$ as given in eq.~\ref{xiassumption}: $$P(\xi \, | \,  \delta_{sky}, G\mu) = \prod_{(i,j)} (p_{i,j})^{\xi_{i,j}}(1-p_{i,j})^{1-\xi_{i,j}}\, .$$
The probability $p_{i,j}$ depends in turn on the parameters $\beta$. We take  ${\bf p} \equiv {\bf p}(\beta)$ to be the probability map
\be\label{pbeta} 
{\bf p}(\beta): \mathbf{R}^{N_{\rm pixel}} \rightarrow [0,1]^{N_{\rm pixel}}
\, .
\ee
It is this probability map that we will present as our prediction maps in section~\ref{resultsStringLocations}.

To find the best parameters $\beta$, we minimize the distance between the parametrised distribution $P_{\beta}(\xi \, | \,  \delta_{sky}, G\mu)$ and the true distribution $P(\xi \, | \,  \delta_{sky}, G\mu)$ using the Kullback-Leibler divergence to measure the distance: 
\bea\label{KLdiv}
D_{KL}(P||P_\beta) &\equiv& 
\sum_{\xi \, \in \,  \mathbf{\Xi}} P(\xi) \log\frac{P(\xi)}{P_\beta(\xi)} 
\\
\label{KLdiv2}
&=& \sum_{\xi \, \in \,  \mathbf{\Xi}} \Big( P(\xi) \log P(\xi) - P(\xi)\log P_\beta(\xi) \Big) \, . 
\eea
It is a measure of the amount of information lost when $P_\beta(\xi)\equiv P_{\beta}(\xi \, | \,  \delta_{sky}, G\mu)$ is used to approximate $P(\xi)\equiv P(\xi \, | \,  \delta_{sky}, G\mu)$. 
The KL divergence is not symmetric with respect to its arguments, hence there is a choice to be made between $D_{KL}(P||P_{\beta})$ and $D_{KL}(P_{\beta}||P)$. We choose the former for multiple reasons.  First, we do not have access to $P$ explicitly (unlike $P_{\beta}$), hence we cannot easily compute $D_{KL}(P_{\beta}||P)$, whereas we can easily approximate the inverse direction with samples from $P$.   Secondly ref.~\cite{KLdirection} also argues that the direction we choose is the optimal one for belief approximation.

The best choice for the parameters $\beta$ is the one that minimizes~\ref{KLdiv}. Since  the first term in~\ref{KLdiv2} is $\beta$ independent, minimizing the Kullback-Leibler divergence is equivalent to minimizing
\be\label{KLdiv3}
\sum_{\xi \, \in \,  \mathbf{\Xi}}\Big( - P(\xi)\log P_\beta(\xi) \Big)\, .
\ee
We can now use simulations to build a dataset of string temperature maps and sky maps.
Given a dataset $\{\xi_i\}$ of maps sampled from this true distribution 
$P(\xi)$ we can approximate ~\ref{KLdiv3} as:
\be\label{KLdiv4}
\sum_{\xi \, \in \,  \mathbf{\Xi}}\Big( - P(\xi)\log P_\beta(\xi) \Big)
\approx \sum_{\xi' \sim P(\xi)} -\log P_\beta(\xi') \, ,
\ee
which we can evaluate and minimize. 

By using~\ref{xiassumption} we can expand~\ref{KLdiv4} even more, 
\bea\label{measure}
&&\log P_\beta(\xi) = \log \prod_{(i,j)}(p_{\beta, (i,j)}(\delta_{sky}))^{\xi_{i,j}}(1-p_{\beta, (i,j)}(\delta_{sky}))^{1-\xi_{i,j}} \nonumber\\
&&= \sum_{(i,j)} \xi_{i,j} \log (p_{\beta, (i,j)}(\delta_{sky})) + (1-\xi_{i,j})\log (1- p_{\beta, (i,j)}(\delta_{sky})) 
\eea
where $p_{\beta, (i,j)}(\delta_{sky})$ is simply the $(i,j)^{th}$ output of $p(\beta)$ from eq.~\ref{pbeta} applied on $\delta_{sky}$. 

Eq.~\ref{measure} represents the function which needs to be minimized by a careful choice of $\beta$ in order to learn to approximate the probability distribution. By choosing a function $p_{\beta} (\delta_{sky})$ which is differentiable with respect to $\beta$, gradient descent optimization algorithms can be used to set the parameters to values which minimize the Kullback-Leibler divergence. 
In particular, we 
parametrise the function $p_{\beta, (i,j)} (\delta_{sky})$ in eq.~\ref{measure} by using a neural network. Neural networks are a class of non-linear function approximators inspired by human brain architecture and are a part of the field of machine learning (for a recent review of the field of neural networks, see \cite{BengioReview}.)
We describe in detail the neural network we use in ref.~\cite{nextpaper}. 
We denote the neural networked optimized values for $\beta$ as $\bar\beta$ and thus we can use eq.~\ref{xiassumption} to write eq.~\ref{PGskyBis} as:
\be\label{PGskyTris}
P(G\mu \, | \, \delta_{sky})  \approx \Big({1\over2}\Big)^{N_{\rm pixel}}\, 
\Bigg(
\frac{P(G\mu)}{P(\delta_{sky})} 
\Bigg)
\Bigg\{
\frac{1}{n} \sum_{k=1}^n  \frac{ P (\delta_{sky} - G\mu\,  \delta_{string}^k ) }{ \prod_{i,j} (p_{\bar\beta})^{\xi^k_{i,j}}(1-p_{\bar\beta})^{1-\xi^k_{i,j}} }
\Bigg\} \, ,
\ee
where the function $p_{\bar\beta}\equiv p_{\bar\beta, (i,j)} (\delta_{sky})$ in the equation above is the probability of success for the string location map $\xi^k$ which determines $\delta_{string}^k$.

\section{Results: Neural Network Predictions for String Locations}
\label{resultsStringLocations}
%
\begin{figure}
\centering
\caption{CMB anisotropy temperature maps 
of $512\times512$ pixels with a resolution of 1 arcminute per pixel.}
\begin{subfigure}[b]{0.48\textwidth}
\includegraphics[width=1.0\textwidth]{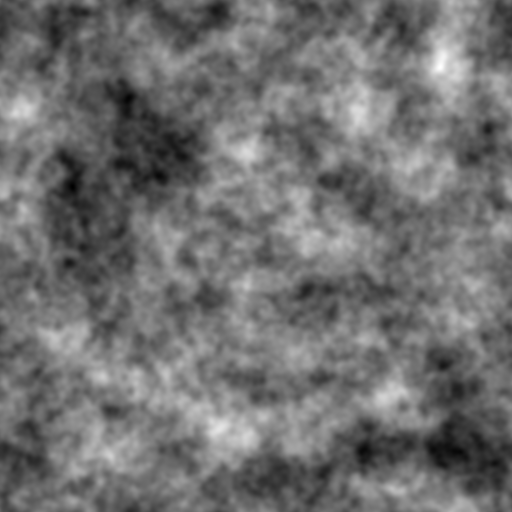}
\caption{The full sky map, $\delta_{sky}$. 
The white and black pixels are $+450\mu$K and $-450\mu$K anisotropies, respectively. Maps with and without strings are indistinguishable by eye. 
}
\label{skymap}
\end{subfigure}
\begin{subfigure}[b]{0.48\textwidth}
\includegraphics[width=1.0\textwidth]{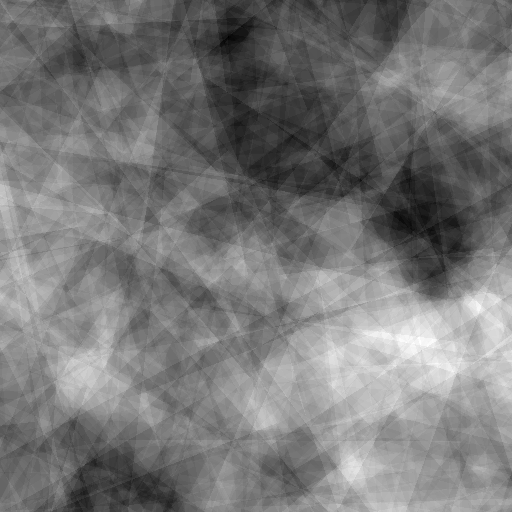}
\caption{String component $\delta_{string}$ to the full sky map.\\  \\ \\}
\label{stringmap}
\end{subfigure}
\end{figure}
We trained the neural network presented in ref.~\cite{nextpaper} on numerically generated CMB temperature maps with and without cosmic strings. This dataset was obtained with the same long string analytical model~\cite{Perivolaropoulos:1992if} used by previous studies of cosmic string detection in CMB maps~\cite{Stewart:2008zq,Danos:2008fq,Hergt:2016xup}. We used the PyTorch environment (pytorch.org) for machine learning and optimization algorithms. Training the model on a Tesla K80 GPU took 12 hours in total. 

We use the analytic long string model~\cite{Perivolaropoulos:1992if} to simulate CMB temperature anisotropy maps with strings. 
The maps were made up of $512\times512$ pixels with a resolution of 1 arcminute per pixel. 
This leads to the sky map show in Fig.~\ref{skymap}. For values of the string tension we study here, $G\mu\leq10^{-7}$, the sky map is indistinguishable by eye from a pure Gaussian fluctuation map (i.e. $G\mu=0$). 
The string temperature component to the full sky map is shown in Fig.~\ref{stringmap} with a $G\mu=1$.
One of the unknown parameters characterizing the scaling solution of strings 
is the number of strings per Hubble volume, $N_H$, which can have a value between 1 and 10. We trained our neural network with a value of $N_H=1$ and this did not impair the predictive power for input maps with larger $N_H$ values. This is an indication that the network is indeed generalizing and not just overfitting.
We also tested the robustness of the network to noise by adding a white noise component. Below we highlight results which confirm the soundness and power of the Bayesian machine learning framework and its neural network implementation, that we have presented here.
Further details of the neural network and more detailed presentation of the results are in~\cite{nextpaper}. 

\begin{figure}
\centering
\begin{subfigure}[b]{0.45\textwidth}
\includegraphics[width=\textwidth]{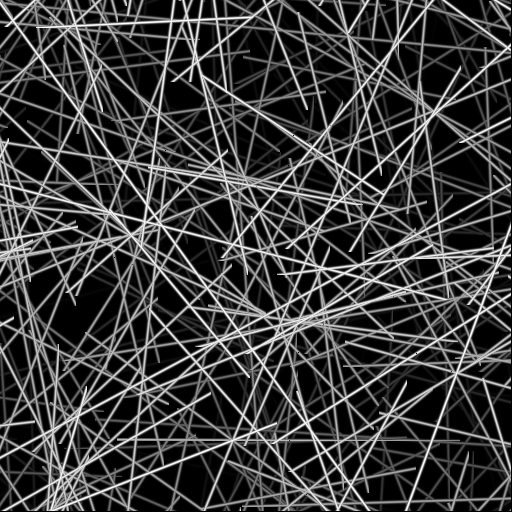}
\caption{Map of strings used in the simulation}
\label{strings}
\end{subfigure}
\begin{subfigure}[b]{0.45\textwidth}
\includegraphics[width=\textwidth]{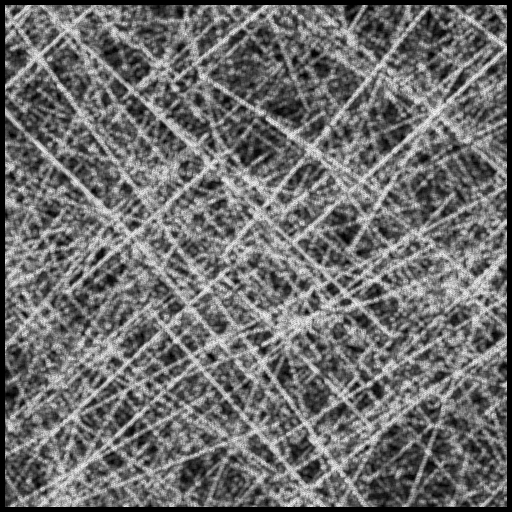}
\caption{Prediction  when $G\mu=10^{-7}$}
\label{prediction1e-7}
\end{subfigure}
\begin{subfigure}[b]{0.45\textwidth}
\includegraphics[width=\textwidth]{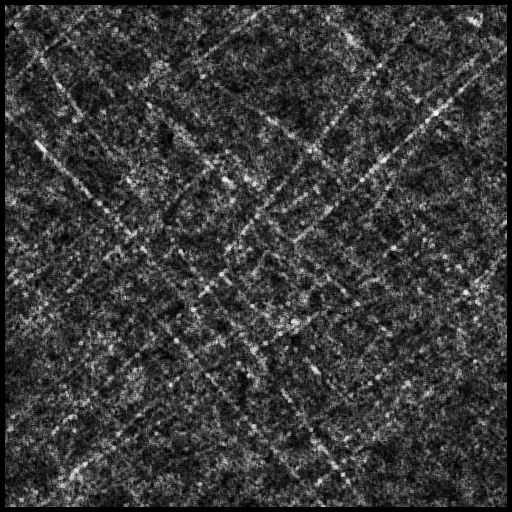}
\caption{Prediction  when $G\mu=10^{-8}$}
\label{prediction1e-8}
\end{subfigure}
\begin{subfigure}[b]{0.45\textwidth}
\includegraphics[width=\textwidth]{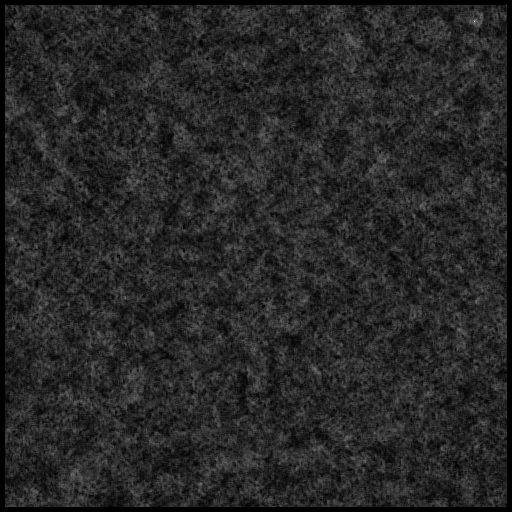}
\caption{Prediction  when $G\mu=5\times10^{-9}$}
\label{prediction5e-9}
\end{subfigure}
\caption{{\bf Neural Network Predictions Without Noise.} 
All the figures correspond to $512\times512$ pixels with a resolution of 1 arcminute per pixel.
In \ref{strings} we show our answer map, i.e. the actual placement of long strings in our patch of sky for $N_H=3$. In \ref{prediction1e-7}, \ref{prediction1e-8}, \ref{prediction5e-9} we show our neural network's prediction of $\xi$ for different value of the string tension with no noise. The shades of grey of the strings in the string answer map correspond to the relative strength of the string's GKS temperature discontinuity. 
The shades of grey in the prediction maps correspond to the probability of a pixel being on a string, with completely black pixels being 0 probability and completely white pixels being probability 1. }
\label{stringpredictions}
\end{figure}
In Fig.~\ref{stringpredictions} we show our neural network predictions for the string location map (see eq.~\ref{pbeta}) using different values for $G\mu$, with $N_H=3$, and no noise. The shades of grey in the string answer map correspond to the relative strength of the string's GKS temperature discontinuity. 
The shades of grey in the prediction maps correspond to the probability of a pixel being on a string. Completely black pixels are probability 0 and completely white pixels are probability 1 of being on a string.  As $G\mu$ tends to zero, the neural network provides less information of whether a pixel is on a string or not and the pixel probabilities tend to the prior $P((i,j)\in string)$ which is given by the number of pixels on strings in the Answer map (Fig.~\ref{strings}) divided by the total number of pixels. Thus as $G\mu$ tends to zero, our prediction map will become more uniformly grey, as \ref{prediction5e-9} shows. 

Looking at Fig.~\ref{stringpredictions}, we see that we can reconstruct string locations by eye at $G\mu\sim 10^{-8}$. 
To compare our results with the wavelet curvelet of ref.~\cite{Hergt:2016xup} look at their figure 3 and 5 for $G\mu=10^{-7}$. Our neural network predictions produces string location maps comparable to their figure 3 and 5 but for a string tension $G\mu$ an order of magnitude lower. 
We now compare in more detail our neural network predictions, with the results obtained by using the Canny algorithm.

\begin{figure}
\centering
\begin{subfigure}[b]{0.45\textwidth}
\includegraphics[width=\textwidth]{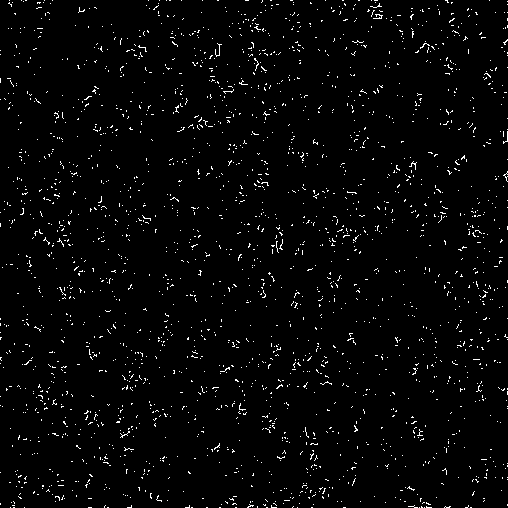}
\caption{No strings, i.e. $G\mu=0$}
\label{canny0}
\end{subfigure}
\begin{subfigure}[b]{0.45\textwidth}
\includegraphics[width=\textwidth]{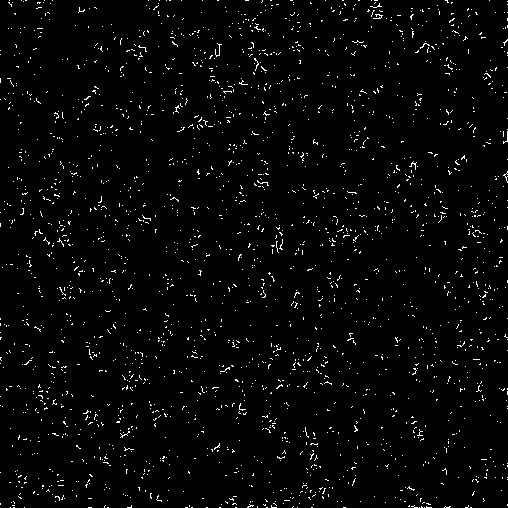}
\caption{$G\mu=10^{-8}$}
\label{canny1e-8}
\end{subfigure}
\begin{subfigure}[b]{0.45\textwidth}
\includegraphics[width=\textwidth]{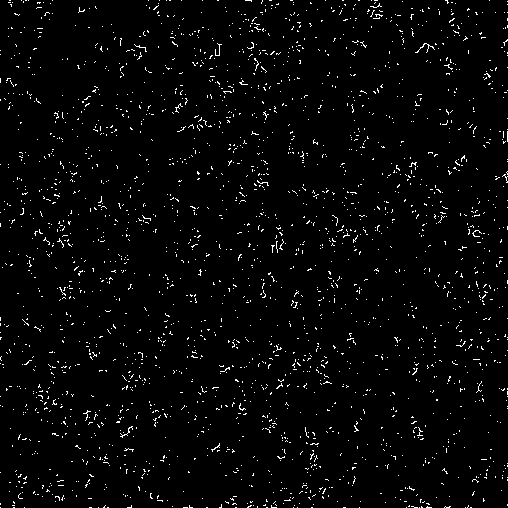}
\caption{$G\mu=10^{-7}$}
\label{canny1e-7}
\end{subfigure}
\begin{subfigure}[b]{0.45\textwidth}
\includegraphics[width=\textwidth]{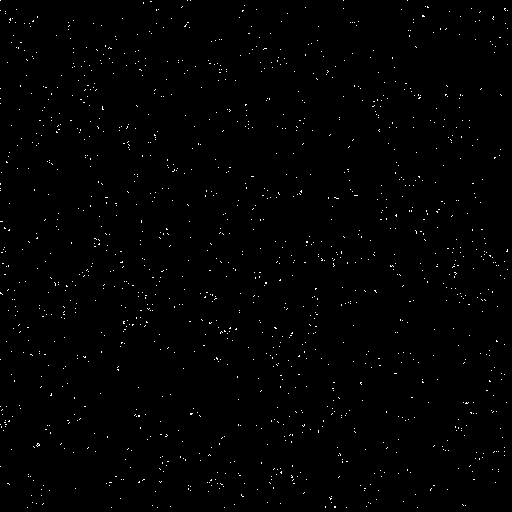}
\caption{Difference between $G\mu=0$ and $10^{-7}$ map.}
\label{diffcanny}
\end{subfigure}
\caption{{\bf Canny Edge Detection Without Noise.}
All the figures correspond to $512\times512$ pixels with a resolution of 1 arcminute per pixel.
In~\ref{canny0} we show the Canny edge map of pure Gaussian fluctuations without strings. Figures~\ref{canny1e-8} and \ref{canny1e-7} show the Canny edge map with $G\mu=10^{-8}$ and $10^{-7}$, respectively. The true string locations for these maps are given in Fig.~\ref{strings}. Fig.~\ref{diffcanny} shows only those edges that appear in the $G\mu=10^{-7}$ map Fig.~\ref{canny1e-7} but not in the no string map Fig.~\ref{canny0}. The edges in Fig.~\ref{diffcanny} occupy 1862 of the $512\times512$ pixels.
}
\label{canny}
\end{figure}
\begin{figure}
\centering
\begin{subfigure}[b]{0.45\textwidth}
\includegraphics[width=\textwidth]{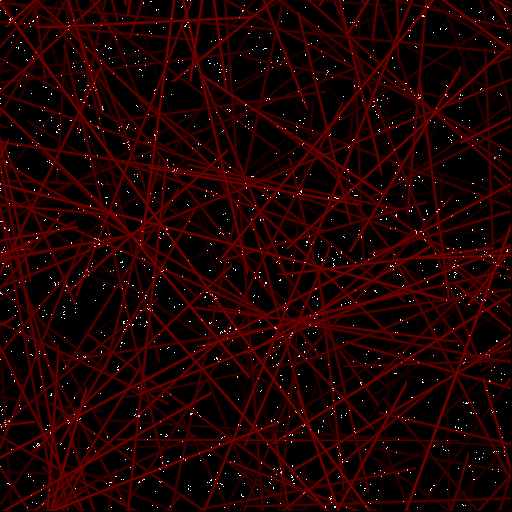}
\caption{Canny string edges for $G\mu=10^{-7}$}
\label{cannyoverlay}
\end{subfigure}
\begin{subfigure}[b]{0.45\textwidth}
\includegraphics[width=\textwidth]{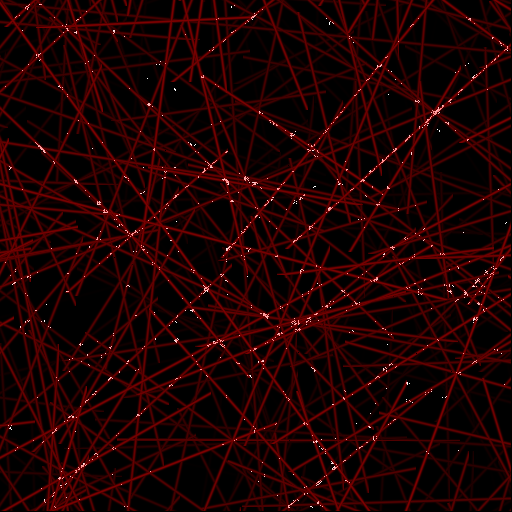}
\caption{NN prediction for $G\mu=10^{-7}$}
\label{prediction1e-7overlay}
\end{subfigure}
\begin{subfigure}[b]{0.45\textwidth}
\includegraphics[width=\textwidth]{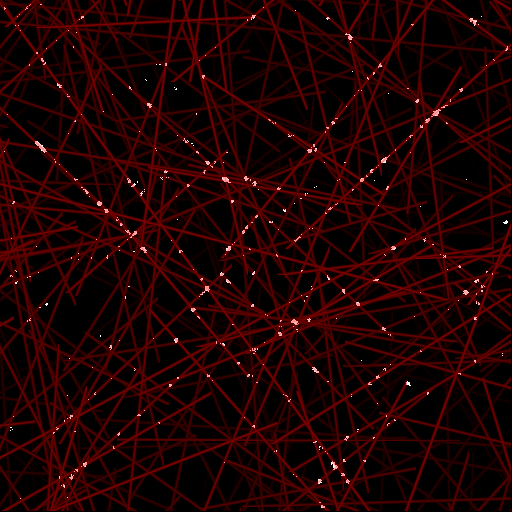}
\caption{NN prediction for $G\mu=10^{-8}$}
\label{prediction1e-8overlay}
\end{subfigure}
\begin{subfigure}[b]{0.45\textwidth}
\includegraphics[width=\textwidth]{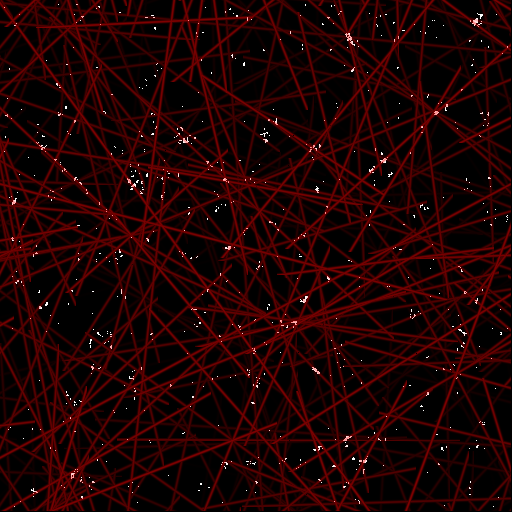}
\caption{NN prediction for $G\mu=5\times10^{-9}$}
\label{prediction5e-9overlay}
\end{subfigure}
\caption{{\bf Overlay of Canny and Neural Network (NN) Predictions with String Locations.} 
All the figures correspond to $512\times512$ pixels with a resolution of 1 arcminute per pixel.
We show the overlay of the actual string locations from Fig.~\ref{strings} in red with the Canny edge map (Fig.~\ref{diffcanny}) and the 1862 brightest pixels from our prediction maps for different string tension (Fig.~\ref{prediction1e-7},\ref{prediction1e-8},\ref{prediction5e-9}) in white.
}
\label{overlay}
\end{figure}
\begin{figure}
\centering
\begin{subfigure}[b]{0.44\textwidth}
\includegraphics[width=\textwidth]{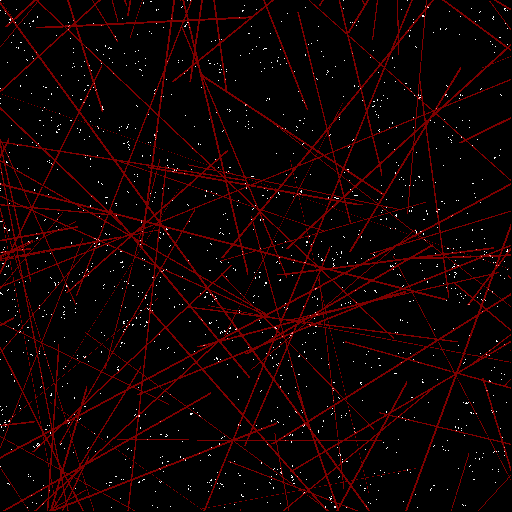}
\caption{Canny string edges for $G\mu=10^{-7}$}
\label{1862canny}
\end{subfigure}
\begin{subfigure}[b]{0.44\textwidth}
\includegraphics[width=\textwidth]{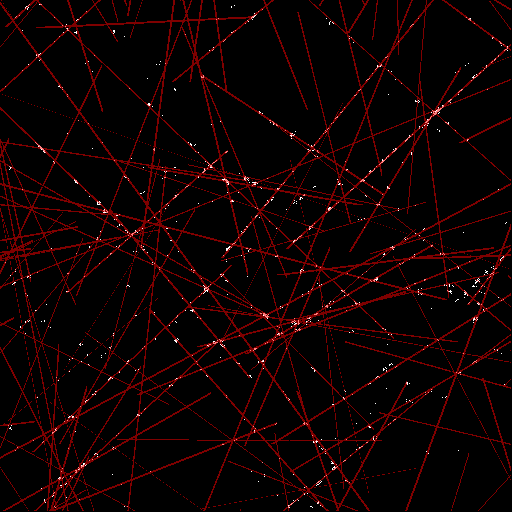}
\caption{NN prediction for $G\mu=10^{-7}$}
\label{1862prediction1e-7}
\end{subfigure}
\begin{subfigure}[b]{0.44\textwidth}
\includegraphics[width=\textwidth]{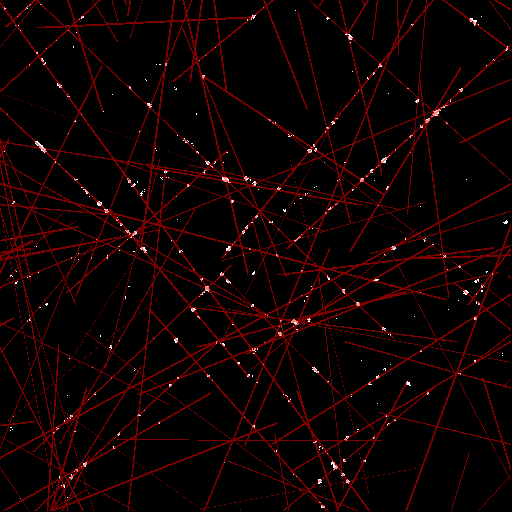}
\caption{NN prediction for $G\mu=10^{-8}$.}
\label{1862prediction1e-8}
\end{subfigure}
\begin{subfigure}[b]{0.44\textwidth}
\includegraphics[width=\textwidth]{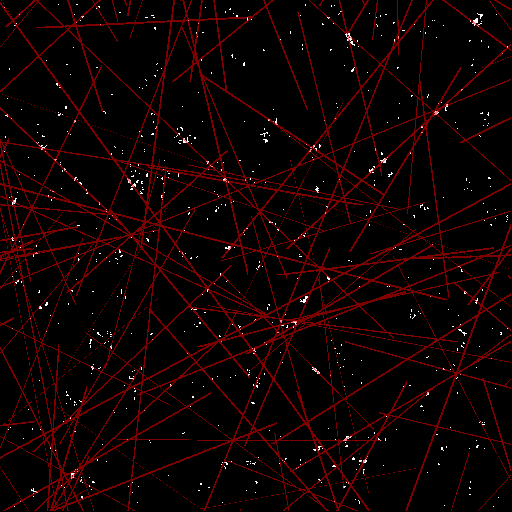}
\caption{NN prediction for $G\mu=5\times10^{-9}$.}
\label{1862prediction5e-9}
\end{subfigure}
\caption{{\bf Overlay of Canny and Neural Network (NN) Predictions with 30,000 brightest String Location Pixels.} 
All the figures correspond to $512\times512$ pixels with a resolution of 1 arcminute per pixel.
We show the overlay of the 30,000 brightest pixels from Fig.~\ref{strings} in red with the Canny edge map (Fig.~\ref{diffcanny}) and the 1862 brightest pixels from our prediction maps with different string tension (Fig.~\ref{prediction1e-7},\ref{prediction1e-8},\ref{prediction5e-9}) in white.
}
\label{fig:true1862}
\end{figure}
\begin{figure}
\centering
\begin{subfigure}[b]{0.44\textwidth}
\includegraphics[width=\textwidth]{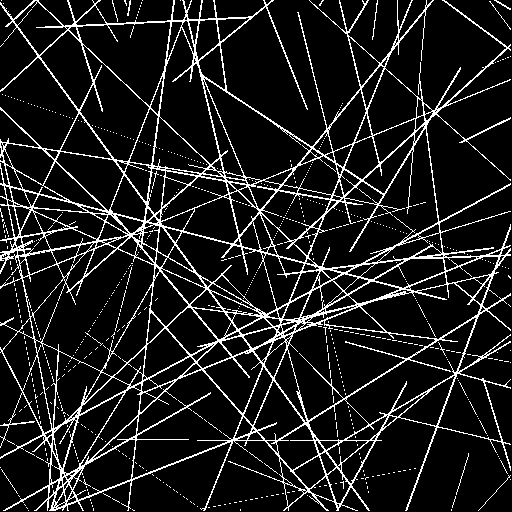}
\caption{The 30,000 brightest pixels from Fig.~\ref{strings}}
\label{1030strings}
\end{subfigure}
\begin{subfigure}[b]{0.44\textwidth}
\includegraphics[width=\textwidth]{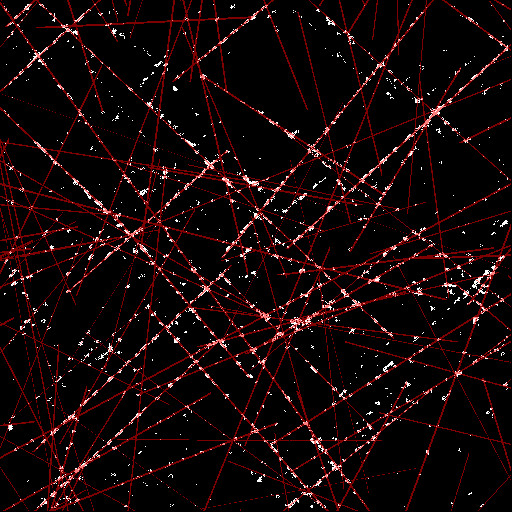}
\caption{NN prediction for $G\mu=10^{-7}$}
\label{1030prediction1e-7}
\end{subfigure}
\begin{subfigure}[b]{0.44\textwidth}
\includegraphics[width=\textwidth]{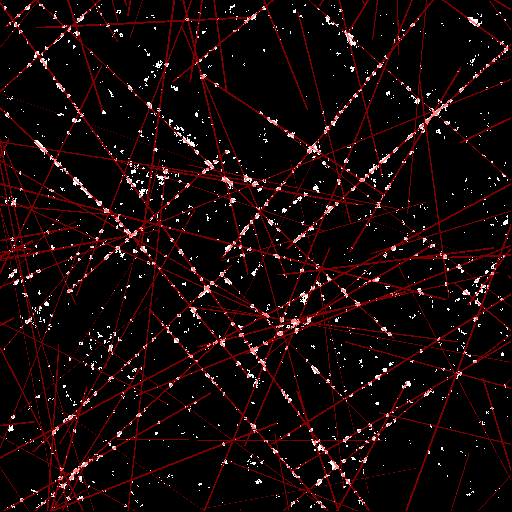}
\caption{NN prediction for $G\mu=10^{-8}$.}
\label{1030prediction1e-8}
\end{subfigure}
\begin{subfigure}[b]{0.44\textwidth}
\includegraphics[width=\textwidth]{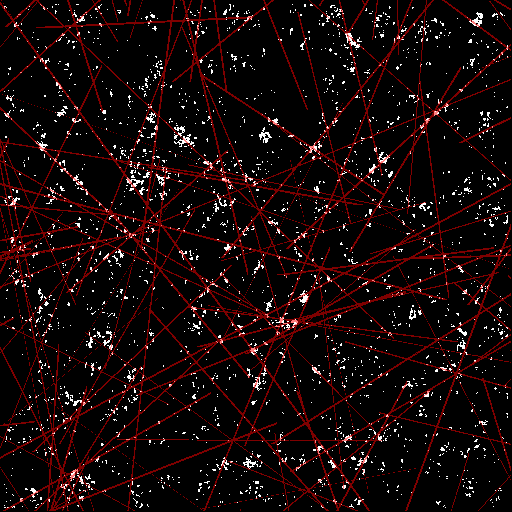}
\caption{NN prediction for $G\mu=5\times10^{-9}$.}
\label{1030prediction5e-9}
\end{subfigure}
\caption{{\bf Overlay of the 10,000 brightest Neural Network Prediction Pixels with 30,000 brightest String Location Pixels.} 
All the figures correspond to $512\times512$ pixels with a resolution of 1 arcminute per pixel.
Fig.~\ref{1030strings} shows the 30,000 brightest pixels from Fig.~\ref{strings} in white.
 We overlay of these same brightest pixels in red over the 10,000 brightest pixels in white from our prediction maps for different string tensions (Fig.~\ref{prediction1e-7},\ref{prediction1e-8},\ref{prediction5e-9}).
}
\label{fig:1030}
\end{figure}
%
\begin{table}
\centering
\caption{
The rate at which string pixel locations are correctly predicted when compared to the 30,000 brightest string location pixels of Fig.~\ref{strings}. For each case given in column 1, we present in column 2 how many of the 1862 or 10,000 pixels (Table~\ref{table:true1862} or~\ref{table:1030}, respectively) correspond to string pixels. Column 3 gives the true positive rate which is the ratio of the column 2 number divided by 1862 or 10,000. The false positive rate is just 1 minus the true positive rate. 
}
\begin{subtable}{1.0\textwidth}
\centering
\caption{
The rate at which string pixel locations are correctly predicted by the 1862 pixels in the Canny edge map and the 1862 brightest pixels in the neural network (NN).
}
	\begin{tabular}{ccc}\hline\hline
	$G\mu$ & No. of pixels corresponding to string pixels & True positive rate\\
	\hline\hline
	Canny: $1 \times 10^{-7}$& 310 & 17\% \\ \hline
	NN: $1 \times 10^{-7}$& 1117 & 60\% \\
	NN: $1 \times 10^{-8}$& 1060 & 57\% \\
	NN: $5 \times10^{-9}$& 423 & 23\% \\
	NN: $1\times10^{-9}$& 243 & 13\% \\
	\\
	\end{tabular}
\label{table:true1862}
\end{subtable}
\begin{subtable}{1.0\textwidth}
\centering
\caption{
The rate at which string pixel locations are correctly predicted by the 10,000 brightest pixels in the neural network.
}
	\begin{tabular}{ccc}\hline
	$G\mu$ & No. of pixels corresponding to string pixels & True positive rate\\
	\hline
	$1 \times 10^{-7}$& 5001 & 50\% \\
	$1 \times 10^{-8}$& 4089 & 41\% \\
	$5 \times10^{-9}$& 1936 & 19\% \\
	$1 \times10^{-9}$& 1205 & 12\% \\
	\end{tabular}
\label{table:1030}
\end{subtable}
\end{table}

We had access to the C code used by~\cite{Amsel:2007ki} and~\cite{Stewart:2008zq} to perform the CMB simulations and then analyzed them with the Canny algorithm. We rewrote significant parts of the code both to generate maps and reconstruct the predictions of the Canny algorithm~\cite{nextpaper}. We found that without noise Canny can only distinguish between strings and no strings  for a  $G\mu\gtrsim 10^{-7}$~\cite{nextpaper} and when noise is included this drops to $G\mu\gtrsim 10^{-6}$. These limits are similar to those obtained in~\cite{Hergt:2016xup}. We present our Canny edge maps in Fig.~\ref{canny}. The pure Gaussian, no string edge map is given in Fig.~\ref{canny0}, and those for string tension $G\mu=10^{-8}$ and $10^{-7}$ in Fig.~\ref{canny1e-8} and Fig.~\ref{canny1e-7}, respectively. Fig.~\ref{diffcanny} shows only those edges that appear in the $G\mu=10^{-7}$ map of Fig.~\ref{canny1e-7} but not in the no string map of Fig.~\ref{canny0}.
The Canny edge maps in Fig.~\ref{canny} are produced with the same procedure described in ref.~\cite{Amsel:2007ki,Stewart:2008zq,Danos:2008fq}.  In particular see figure 5 in~\cite{Stewart:2008zq} and figures 13, 14 and 15 in~\cite{Danos:2008fq}. 

As explained in~\cite{Amsel:2007ki,Stewart:2008zq,Danos:2008fq}, Canny is used to distinguish between maps with and without strings by looking for an excess number of short edges (a few pixels or less) over the entire map. This can be noted by looking at Fig.~\ref{diffcanny} where we show those excess edges that appear in a CMB temperature map with $G\mu=10^{-7}$. There are hundreds of short edges comprising of 1862 pixels out of the entire $512\times512$ pixel map. 
This is interpreted as the long edges due to strings being disrupted by the Gaussian noise. However the extra short edges found by Canny do not necessarily correspond to string locations.  

To visually compare how well Canny and our neural network pick out the brightest strings, we present in Figures~\ref{overlay},~\ref{fig:true1862}, and~\ref{fig:1030} an overlay of their predictions with the actual string locations taken from Fig.~\ref{strings}. In Figure~\ref{overlay} the 1862 excess Canny pixels and 1862 brightest neural network pixels are compared to all 156,137 string pixels in the map whereas in Figure~\ref{fig:true1862} they are only compared to the 30,000 brightest string pixels. We chose the 1862 brightest pixels in our prediction maps~\ref{prediction1e-7}, \ref{prediction1e-8}, \ref{prediction5e-9} since the Canny edge map contains only 1862 nonzero pixels due to strings.  Finally Figure~\ref{fig:1030} compares the 10,000 brightest neural network string prediction pixels to the 30,000 brightest string pixels.

In Fig.~\ref{overlay} we have overlaid Fig.~\ref{strings} on Fig.~\ref{diffcanny},\ref{prediction1e-7},\ref{prediction1e-8},\ref{prediction5e-9}.
There we can easily appreciate that the neural network is doing a very good job in finding string locations for string tensions of $G\mu=10^{-7}$ and $10^{-8}$ and less well for $G\mu=5\times10^{-9}$, whereas for Canny almost all of the extra edges do not fall on strings. 

Because strings have varying velocity, not all the 156,137 string pixels give an equally bright signal. Furthermore, these 156 thousand pixels occupy 60\% of the $512\times512$ pixels in the map. For these reasons a comparison with a reduced number of the brightest string pixels would give a more stringent test of the success of locating strings. Hence we take the 30,000 brightest pixels in the $512\times512$ string location map and show the visual overlap in Fig.~\ref{fig:true1862}.
We also calculate the true positive rate of predicted string locations with Canny and our neural network predictions maps for these 1862 brightest prediction pixels compared to the 30,000 pixels. These results are presented in Table~\ref{table:true1862}. From the table we see that the true positive rates for pixels on strings given by the neural network predictions are much higher than Canny.

Finally, whereas Canny only offers 1862 pixels for comparison, our neural network is not limited to that number. We could for example look at the 10,000 brightest pixels of our prediction maps and compare these to the 30,000 brightest pixels of the string location map. We present these results in Table~\ref{table:1030} and show the visual overlap in Fig.~\ref{fig:1030}.

\section{Results: Neural Network Predictions for the String Tension}
\label{resultsGmuPosterior}
%
\begin{figure}
\centering
\includegraphics[width=\textwidth]{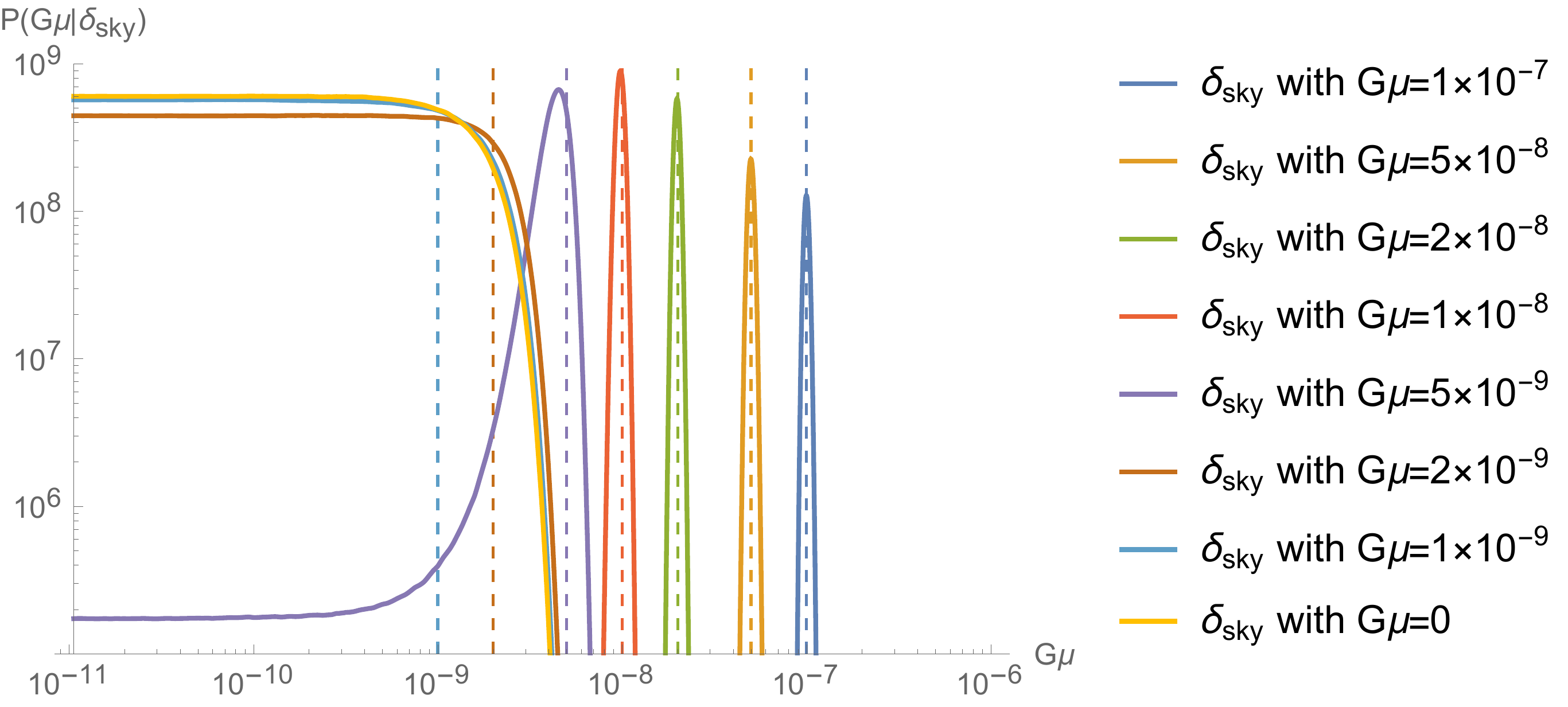}
\caption{The posterior probability $P(G\mu\, | \, \delta_{\rm sky})$ versus the value of $G\mu$ used to simulate the sky map $\delta_{\rm sky}$. Both the values of the posterior probability and of $G\mu$ are on a log scale. The vertical dashed lines are, from left to right, at:
$G\mu=10^{-9}, 2\times10^{-9}, 5\times10^{-9},10^{-8}, 2\times10^{-8}, 5\times10^{-8}, 10^{-7}
$.
}
\label{fig:PGmu}
\end{figure}
\begin{figure}
\centering
\includegraphics[width=\textwidth]{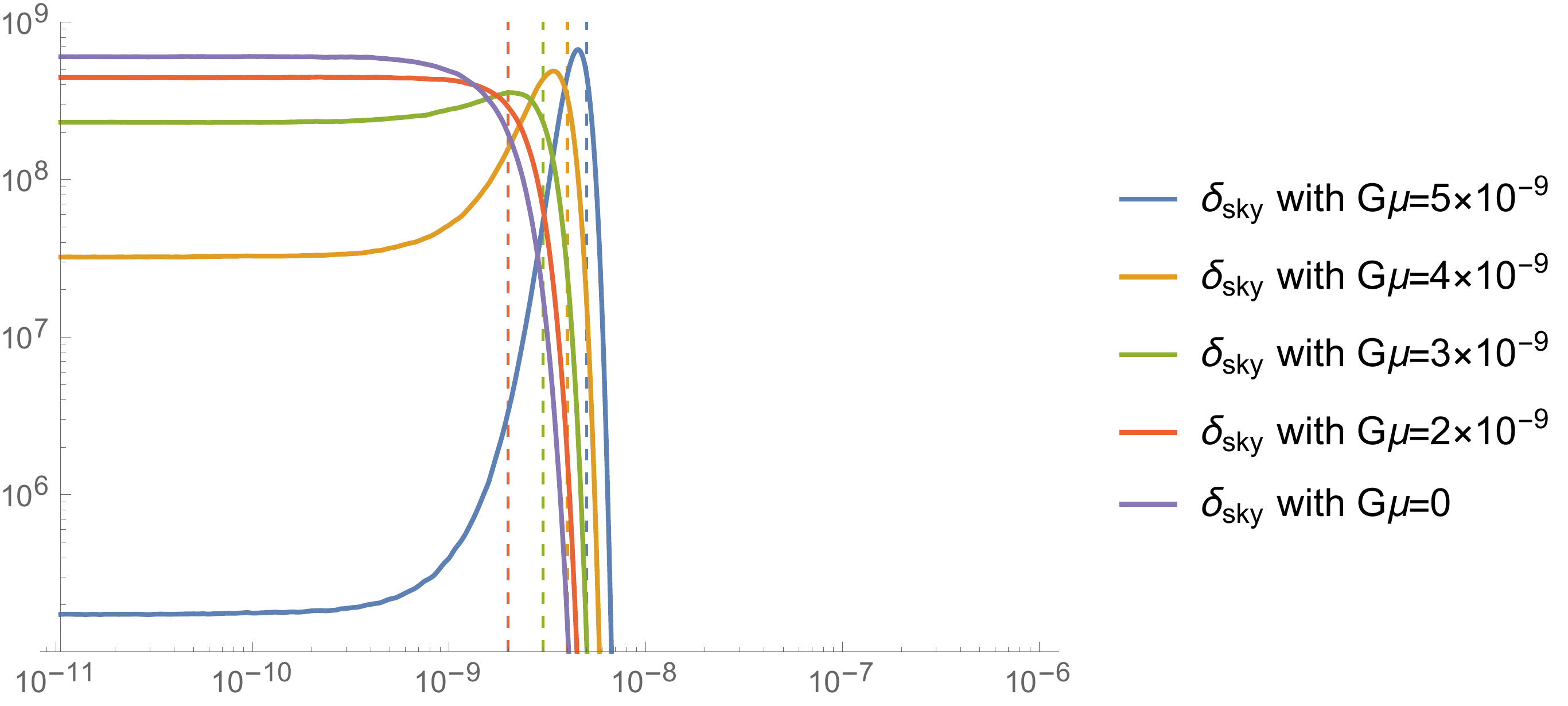}
\caption{The posterior probability $P(G\mu\, | \, \delta_{\rm sky})$ versus the value of $G\mu$ used to simulate the sky map $\delta_{\rm sky}$. Below string tensions of $5\times10^{-9}$ the posterior probability curves spread out so much so that by $G\mu=2\times10^{-9}$ there is no more discernable peak, it simply becomes flat down to zero string tensions. Both the values of the posterior probability and of $G\mu$ are on a log scale. The vertical dashed lines are, from left to right, at:
$2\times10^{-9}, 3\times10^{-9},4\times10^{-9},5\times10^{-9}$.
}
\label{becomesflat}
\end{figure}

Through eq.~\ref{PGskyTris} we can use the prediction maps to provide a probability estimate of the string tension. This leads to interesting constraints even when the maps used do not have strings that are not readily visible, as for example the prediction map in Fig.~\ref{prediction5e-9} with $G\mu=5\times10^{-9}$. 
As our prior, $P(G\mu)$ in eq.~\ref{PGskyTris}, we take an exponentially decaying function with a decay constant chosen so that  95\% of the probability is below a $G\mu$ of $10^{-7}$. 
In Fig.~\ref{fig:PGmu} and Fig.~\ref{becomesflat} we show the probability $P(G\mu\, | \, \delta_{\rm sky})$ versus the value of $G\mu$ used to simulate the sky map $\delta_{\rm sky}$ for different values of the string tension in the sky map. When these probability distributions have a peak we can consider the range of $G\mu$ values about this peak.  
Below string tensions of $4\times10^{-9}$ the probability distribution begins to spread and the peak is less and less discernable. 
For a string tension of $4\times10^{-9}$, the peak is at $G\mu=3.3\times10^{-9}$ and 95\% of the curve's area is between $1.2 \times10^{-9}$ and $5.3\times10^{-9}$. In other words the neural network is predicting that there is a 0.95 probability that the string tension in the sky map is between these two values. 
For a string tension of $5\times10^{-9}$, the peak is at 
$G\mu=4.5\times10^{-9}$ and 95\% of the curve's area is between $3.2 \times10^{-9}$ and $5.7\times10^{-9}$.
For string tensions above $10^{-8}$ this posterior probability is even more sharply peaked about the string tension used to simulate the map. At $G\mu=10^{-8}$, for example, 99\% of the curve's area is between $0.86 \times10^{-8}$ and $1.1 \times10^{-8}$ and at $G\mu=10^{-7}$, 99.9\% of the area is between $0.92\times10^{-7}$ and $1.1\times10^{-7}$.

Finally, we can consider posterior probability curve of $G\mu$ when analyzing a map that does not contain strings. This is the $G\mu=0$ curve in figures~\ref{fig:PGmu} and~\ref{becomesflat}.
For this curve 95\% of the area is below $2.3\times10^{-9}$. Thus there is a 0.95 probability that $G\mu$ is below this value. 
Though not exactly equivalent this  upper limit 
is comparable to the upper limits quoted at a 95\% confidence level, given in the frequentist approach. 

\section{Conclusions}
\label{theend}
We have presented a Bayesian interpretation of cosmic string detection 
and proposed a general machine learning framework.
We have implemented this approach with a convolution neural network trained on simulations of CMB temperature anisotropy maps with and without strings and used it to estimated string locations on a CMB sky temperature map. We have shown a connection between these estimated string locations, called the prediction maps, and the posterior probability of the string tension $G\mu$ given the sky map. 

We have presented these prediction maps 
in section~\ref{resultsStringLocations}
and explained their advantage and improved accuracy compared to the Canny algorithm.  Furthermore strings are visible by eye in the prediction maps for string tensions below $G\mu=10^{-8}$.  We showed that the Canny algorithm~\cite{Amsel:2007ki,Stewart:2008zq,Danos:2008fq} does not provide an accurate location of strings for $G\mu=10^{-7}$. And the wavelet curvlet approach of ref.~\cite{Hergt:2016xup} produces maps where strings are visible by eye only for string tensions above $G\mu=10^{-7}$. 

In section~\ref{resultsGmuPosterior} we showed how the posterior probability formula in eq.~\ref{PGskyTris} allows us to determine the string tension accurately for sky maps with $G\mu>4\times10^{-9}$, even though string locations are not readily visible on such 
predictions 
maps. 
Furthermore when analyzing maps that contain no strings, we are able to establish that the value of the string tension is  $G\mu<2.3\times10^{-9}$ with a probability of 0.95. eq.~\ref{PGskyTris} provides a direct calculation of the posterior probability $P(G\mu\, | \, \delta_{\rm sky})$ using the prediction map ${\bf p}(\beta)$ (see eq.~\ref{pbeta}), regardless of how ${\bf p}(\beta)$ is obtained, whether it be a neural network or not. 

The results from this neural network are extremely promising. The true test will be to successfully extend the network so that it performs well in the presence of noise.

\acknowledgments
We would like to acknowledge the support of the Fonds de recherche du Qu\'ebec--Nature et technologies (FRQNT) Programme de recherche pour les enseignants de coll\`ege.  This research was enabled in part by support provided by Calcul Qu\'ebec (www.calculquebec.ca) and Compute Canada (www.computecanada.ca).

\end{document}